# Thermal Diffusivity in Metallic Nanowire Aerogels

Madeline A. Maben,[1] WLNC Namila Liyanage,[1] Cameron S. Jorgensen,[1] Erin Marlowe,[2] Kai Liu,[2] Gerd Duscher,[1] Dustin A. Gilbert[1†]
Materials Science and Engineering, University of Tennessee, Knoxville, TN
Physics Department, Georgetown University, Washington D.C.

†dagilbert@utk.edu

**Abstract**

Metallic nanowire aerogels combine the functional properties of metals with the high porosity of aerogels. Here, Ag and Cu nanowire aerogels with densities of 20–500 mg/cm$^3$ were synthesized by freeze casting and investigated using lock-in thermography during sequential mechanical compaction. Thermal transport depended strongly on nanowire diameter, with 30 nm Ag nanowires exhibiting substantially lower thermal diffusivity than 60 nm wires, likely due to enhanced surface scattering as the wire diameter approaches the electron mean free path in Ag. Surprisingly, increasing the aerogel density through mechanical compaction reduced the calculated thermal diffusivity by nearly 300×, opposite the conventional expectation that denser materials transport heat more efficiently. Further analysis showed that this decrease is dominated by the thickness term in the thermal diffusivity formalism, while the fitted transport coefficient changes comparatively little. The characteristic thermal diffusion timescale decreased only modestly during compaction, indicating that densification primarily collapses pore volume without fundamentally restructuring the conductive nanowire network. This behavior resembles topology-preserving densification, where large geometric compression produces comparatively small changes in the effective percolative transport pathway. These results show that nanowire aerogels cannot always be treated as homogeneous bulk media, because their macroscopic geometry and conductive network evolve differently during compaction.

## Introduction

Metallic nanowire aerogels are an emerging class of nanostructured material which has the potential to combine the high toughness and various functional properties present in metals with the low-density, high surface area to volume qualities of aerogels.[1-3] A variety of advances to critical technologies have the potential to emerge from these metamaterials including advanced batteries, hydrogen storage,[4] catalysts, bioactive filters,[5] electromagnetic shielding,[6] and ultra-light structural materials.[7] Understanding the physical properties of these materials can be quite complex due to the combined effects of nanoscale confinement of the nanowires, ambiguous junctions between nanowires, and microstructural effects arising from the fabrication process.

Nanostructured materials have been investigated for their scaled material properties, including electrical and structural properties.[8-10] Thermal diffusion and conductivity through the nanowire network is particularly challenging to conceptually analyze.[11,12] Thermal conduction in-particular is determined by the multiscale structure of the material, with the heat moving through the nanowires, which have a combination of bulk-like transport and nanoscale confinement effects, then a variable number of point-like inter-wire contacts. Both the surface and the junctions are potentially oxidized, enhancing these effects.

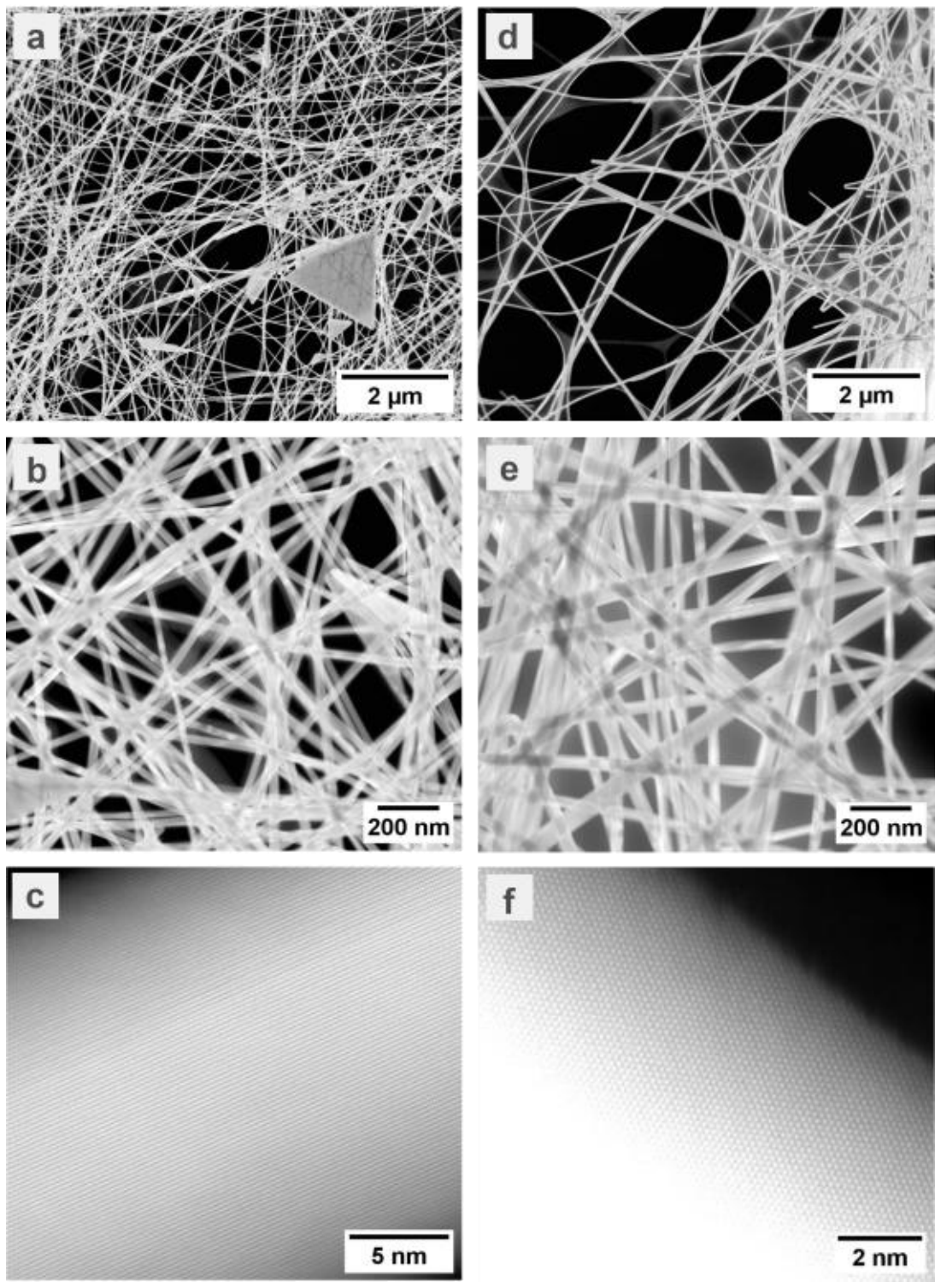

**Figure 1** TEM dark-field (DF) images of A30 (a, b) and A60 (d, e) nanowires at low and high magnification, respectively. High-angle annular dark-field (HAADF) images of A30 (c) and A60 (f) nanowires showing atomic resolution along the [111] direction.

Finally, in these highly porous materials, all three modes of heat transport exist: solid-phase conduction through the wire network, convection through the porous matrix and radiation. Altogether, it is clear that thermal conductivity through these materials is complex and highly dependent on their structure.

While conventional aerogels are made of silica nanowires or carbon nanotubes,[13,14] metallic aerogels present a distinct new direction.[2] The metals carry heat predominantly in their electrons, with full 3D mobility throughout the material. Metals also have much higher ductility compared to the silica or carbon, allowing the structure to have appreciable ductility compared to the more-brittle conventional aerogels, affording new cold-working mechanical processing mechanisms to be employed into their design. Finally, the metal nanowires can be welded, forming a contiguous network and adding substantial strength to the structure.

In the current work, we report on the thermal diffusivity of ultra low-density (20–500 mg/cm$^3$) Ag and Cu nanowire aerogels during post-processing mechanical compaction. In both systems, sequential compaction increased the effective density by more than 30×. Surprisingly, the measured thermal diffusivity decreased by nearly 300×, opposite the conventional expectation that denser materials transport heat more efficiently. Further analysis revealed that this apparent decrease is dominated by the thickness dependence in the lock-in thermography formalism, while the fitted transport coefficient changes comparatively little (<50%). This suggests that compaction primarily collapses pore volume rather than substantially altering the conductive nanowire network. Although the nanowires are brought physically closer together during densification, the dominant heat transport likely still proceeds through nearly the same sequence of nanowires and inter-wire junctions. As a result, the effective percolative transport pathway through the aerogel changes comparatively little despite large geometric compression. Altogether, these results demonstrate a nanostructured material system that can be mechanically compacted by more than 30× while producing comparatively modest changes in the underlying thermal transport through the nanowire network.

## Methods

Silver nanowires with nominal dimensions of (A30) 30 nm diameter × 30 μm length and (A60) 60 nm diameter × 45 μm length were purchased from Novarials and were prepared using a modified polyol reduction process.[15] These diameters were chosen to investigate the role of confinement and surface scattering, as the electron mean free path in silver is 53 nm, falling between the two sizes. Polycrystalline copper nanowires were prepared by electrodeposition into 60 μm thick anodized alumina filter membranes with gold backing from an aqueous solution of $Cu_2SO_4$.[5] The gold was etched using a solution of $KI+I_2$, then the alumina dissolved using a saturated KOH solution. Finally, the Cu nanowires were transferred to deionized water using a series of concentration – decanting – dilution steps. In both systems the aqueous nanowire solutions were concentrated to a slurry with a density of 20-50 mg nanowires/cm$^3$, then suspended by vigorous shaking and/or ultrasonication, then rapidly frozen by submersion in liquid nitrogen. The resulting pellet was then freeze-dried over a 24-hour period. During the freeze-drying process the water is sublimated from the structure; without passing through the liquid phase the nanowires are left intact in their 3D configuration. Micrographs of the as-prepared Ag nanowires were captured using scanning and scanning transmission electron microscopy (SEM and STEM) and high-angle annular dark-field (HAADF) imaging, **Figure 1**. Images of the aerogel structure were captured using SEM in secondary-electron yield contrast mode, **Figures 2** and **3** for A30 and A60, respectively.

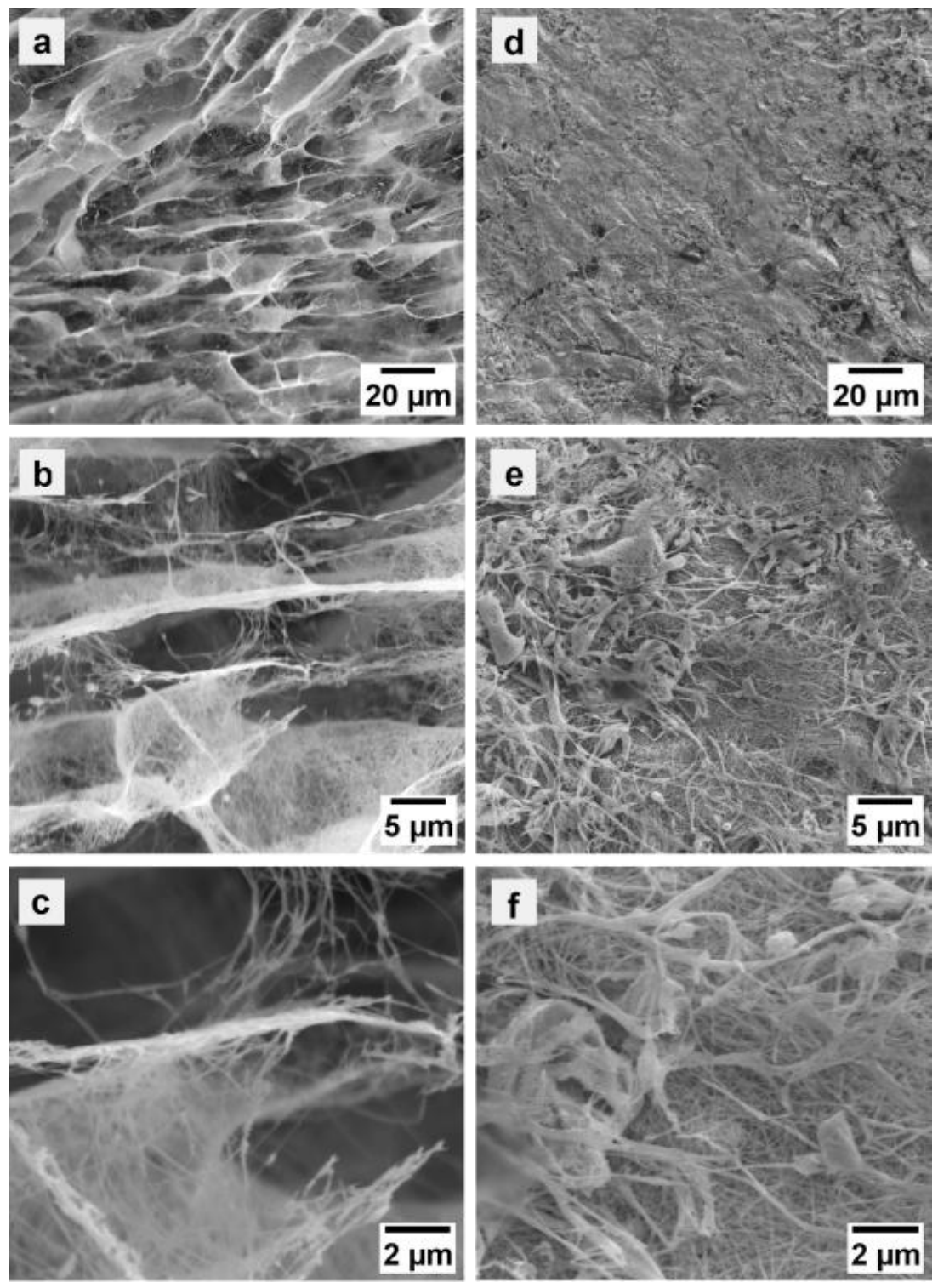

**Figure 2** Plan-view SEM images of as grown A30 aerogel (a, b, c) and A30 aerogel compacted to 0.2 mm (d, e, f).

The as-prepared nanowire aerogels were between 20-50 mg nanowires/cm$^3$. Next, the thermal diffusivity was measured using a laser-based lock-in thermography technique.[16,17] In these measurements, the samples are placed on a glass microscope slide, then the bottom-side is heated with a 1 Watt, 405 nm diode laser; the laser intensity is modulated following a Sine-wave pattern. The time-dependent temperature on the top-side of the sample is measured using an infrared camera. The hottest pixel on the surface of the film is identified and used to identify the laser position. To reduce noise, temperatures were measured by averaging the hottest pixel and 8 nearest neighbors. The time-domain temperature and laser power were captured and treaded as a lock-in signal, capturing the temperature on the top surface and the heating power versus time. Since the heating laser follows a Sine wave, the top-surface temperature also follows a Sine-wave, which is phase-shifted due to the time delay as the heat wave moves through the sample. These two Sine-wave patterns were then fit and the phase delay calculated. Plotting the phase delay versus temperature follows a square-root profile, with a coefficient "b". Using this coefficient and the sample thickness the thermal diffusivity, $\alpha$, is calculated $\alpha = \frac{d^2\pi}{b^2}$.[16] Typically, the diffusivity can be transformed to a thermal conductivity, however we approach this conversion with caution here because the infrared camera captures only the dynamic heat transport through the solid nanowire network, i.e. the conduction, it does not capture convection, for example. Thus, the following results are the thermal diffusivity by conduction through the nanowire matrix. Thermal measurements were performed in air, but tests were also performed in vacuum and no appreciable difference was observed. Measurements were performed at 5 frequencies per-density to determine the diffusivity. After these measurements were completed, the sample was compressed using a series of 3D printed flat-bottomed cups with calibrated depths; in these nanowires aerogels, the compaction is almost entirely irreversible (plastic) deformation. For comparison, thermal diffusivity measurements were also performed on Cu nanowire aerogels that were prepared at 5 wt.%, 10 wt.% and 15 wt.%.

## Results

*Nanowire Aerogel Structures*

Micrographs of the as-received Ag nanowires were dispersed on a lacy carbon grid and imaged, as shown in Fig. 1. The nanowires are shown to be highly-ordered single crystal silver with a <111> growth axis, as is typical for this type of synthesis.[15] Energy dispersive spectroscopy (EDS) confirmed the wires were silver with other contamination below detectable limits. Using line-traces, the nanowire diameters in the A30 samples were confirmed to be 30 ± 5 nm, and the A60 sample to be 70 ± 10 nm; it was not possible to determine the lengths of the nanowires.

Micrographs of the as-prepared aerogels from the A30 wires are shown in Fig. 2. These aerogels are distinct compared to a vast majority of previous works,[18] including our own on Cu[19], Pd[4], or high entropy alloys[17]: previous systems show the nanowires forming random 3D architectures, these nanowires form a veil structure. Notably, there are many nanowires in the plane of the mat, but comparably few wires connecting between the mats, complicating the heat flow through the aerogel. This microstructure emerges in high-density nanowire solutions where the aspect ratio of the nanowires is large: Brownian motion weaves the nanowires into jammed mats, locking the nanowires together and preventing its dissolution. A handful of publications have reported similar structures in Ag nanowires previously;[10,20-23] this distinct structure is likely prevalent in Ag nanowires in-particular because it is relatively simple to prepare large volumes of high-aspect-ratio Ag nanowires by wet-chemical synthesis compared to many other metals.[15]

*Thermography Measurements*

Lock-in thermography measurements were performed on the A30 and A60 nanowire aerogels with initial densities of 24 mg/cm$^3$ and 22 mg/cm$^3$, respectively; these densities are ≈0.2% bulk values. The fitted "b" parameter, which is the coefficient that correlates the phase shift of the topside temperature with respect to the laser, is shown in **Figure 4a**. From these parameters, the thermal diffusivity is calculated, Figure 4b. The thermal diffusivity was calculated to be 0.12 mm$^2$/s and 1 mm$^2$/s for A30 and A60, respectively. For the A30 sample the diffusivity is ≈0.07% of the bulk value of 165 mm$^2$/s, decreasing faster than the density; for the A60 sample the diffusivity is ≈0.6% of the bulk value, scaling slower than the density. Notably, these values are smaller than that of air (≈16 mm$^2$/s) emphasizing the extremely weak solid-phase conduction through the nanowire network.[12] However, the lock-in thermography measurement primarily probes the dynamic thermal response of the solid nanowire scaffold rather than convective heat transport through the pore volume.

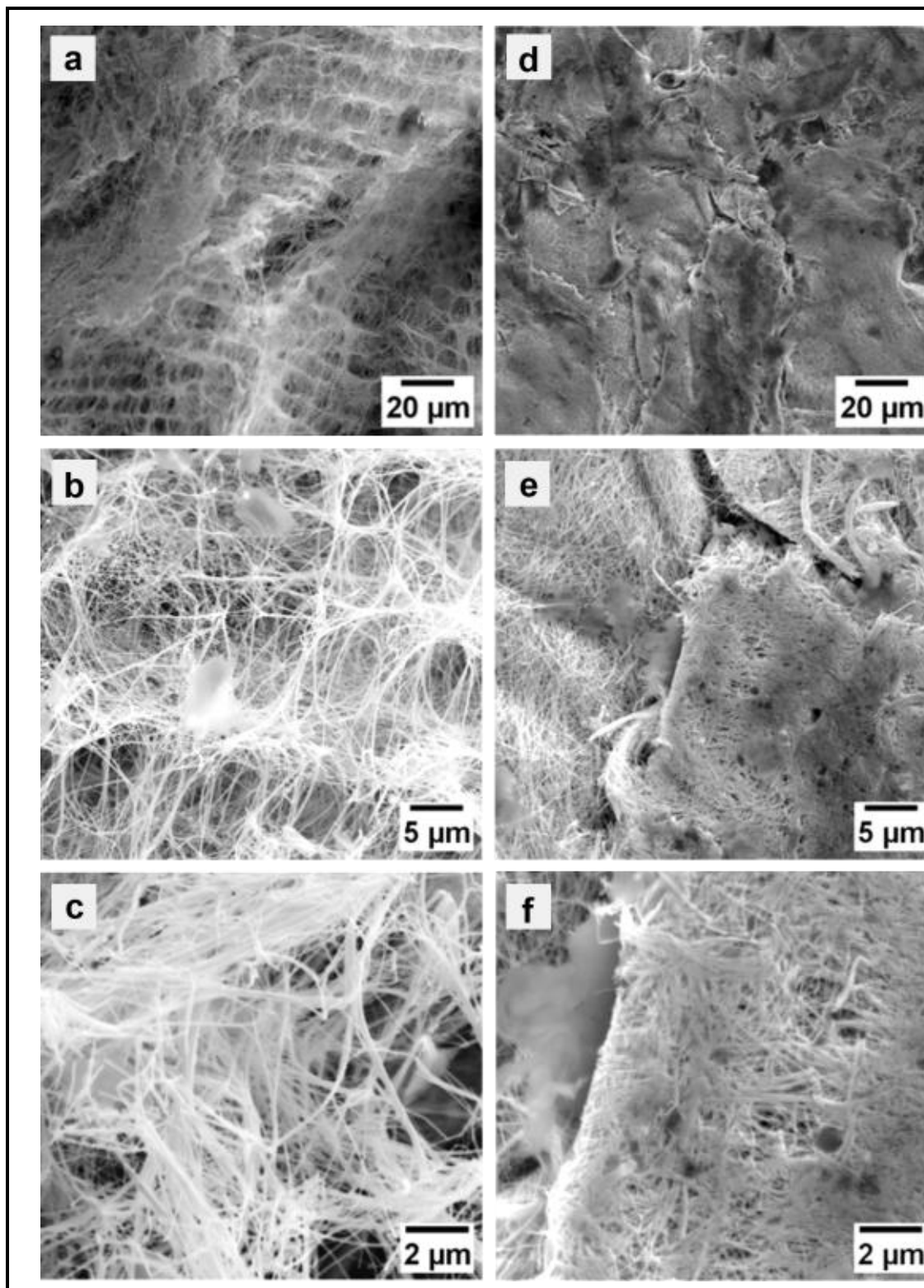

**Figure 3** Plan-view SEM images of as grown A60 aerogel (a, b, c) and A60 aerogel compacted to 0.2 mm (d, e, f).

As the aerogels are compacted the thermal diffusivity decreases by 99.7% for both A30 and A60. This was exactly contrary to the general expectation that increasing the density of a material tends to

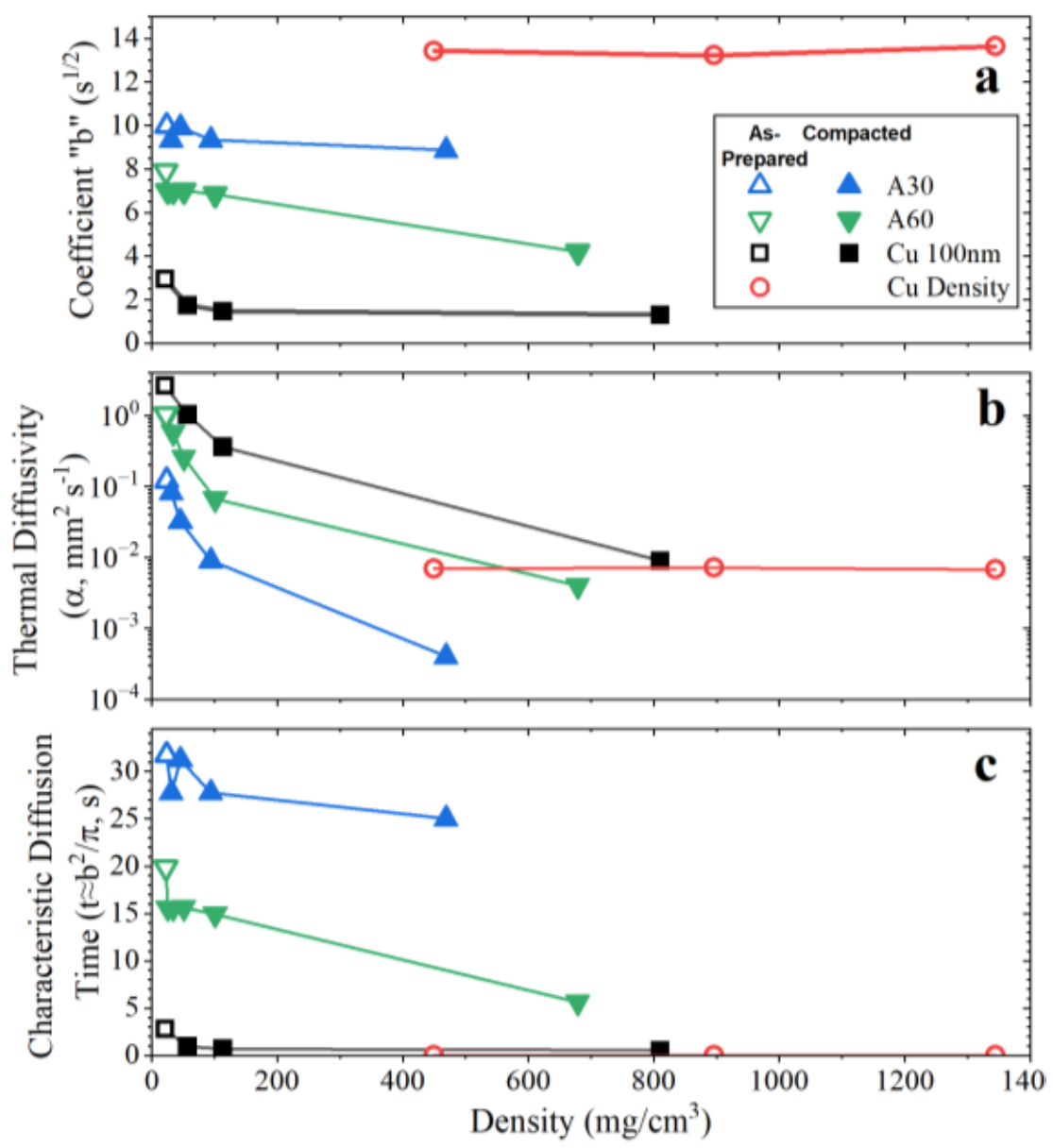

**Figure 4** (a) Fitted "b" parameter describing the frequency-dependence of the phase shift and (b) calculated thermal diffusivity, versus density, (c) calculated thermal transit time.

increase its thermal diffusivity. The diffusivity of the fully compacted A30 and A60 aerogels are 0.0004 $mm^2/s$ and 0.004 $mm^2/s$, respectively. Mechanistically, this means that heat travels through the compacted aerogel 40,000× slower than air, and >400,000× slower than bulk silver.

For comparison, we prepared Cu nanowire aerogels and performed a similar compaction sequence. The Cu aerogel initially had a density of 21 $mg/cm^3$ and thermal diffusivity of 2.6 $mm^2/s$, similar to the A60 sample. As the sample is compacted the same decrease in thermal diffusivity occurs (99.7%), decreasing to a value of 0.009 $mm^2/s$, ≈1,700× slower than air and ≈12,000× slower than bulk copper.

As a final comparison, three samples of Cu nanowire aerogels were measured that were prepared with densities of 45 $mg/cm^3$, 90 $mg/cm^3$, and 135 $mg/cm^3$. These samples were not compacted, separating the roles of the compaction and density. The thermal diffusivity was measured to be 0.007 $mm^2/s$ for each sample – independent of density.

Values for all of the measured aerogels, including the nanowire structure and measured thermal properties, are tabulated in Table 1. While we discourage the comparison to thermal conductivity, due to the nanostructured nature of the material and parallel conduction pathways, we have provided the calculated values in the table as-well.

**Table 1** Structural parameter of the nanowires and nanowire aerogel, fitted diffusivity coefficient, calculated thermal diffusivity and thermal conductivity.

| **Density ($mg\ cm^{-3}$) [Material]** | **Aerogel Thickness (mm)** | **Nanowire Diameter (nm)** | **Fitted coefficient** | **Diffusivity ($mm^2\ s^{-1}$)** | **Conductivity ($mW\ m^{-1}\ K^{-1}$)** |
|---|---|---|---|---|---|
| 21 [Cu]* | 2.7 | 100 | 2.95 | 2.632 | 21.28 |
| 57 [Cu] | 1.0 | 100 | 1.74 | 1.038 | 22.77 |
| 113 [Cu] | 0.5 | 100 | 1.47 | 0.363 | 15.81 |
| 810 [Cu] | 0.07 | 100 | 1.33 | 0.009 | 2.71 |
| 22.5 [Ag, A60]* | 4.53 | 60 | 7.9 | 1.033 | 5.46 |
| 25.5 [Ag, A60] | 4.0 | 60 | 7.0 | 1.026 | 6.15 |
| 34 [Ag, A60] | 3.0 | 60 | 7.0 | 0.577 | 4.61 |
| 51 [Ag, A60] | 2.0 | 60 | 7.0 | 0.256 | 3.07 |
| 101 [Ag, A60] | 1.0 | 60 | 6.85 | 0.067 | 1.60 |
| 679 [Ag, A60] | 0.15 | 60 | 4.0 | 0.004 | 0.71 |
| 24 [Ag, A30]* | 2.0 | 30 | 10.2 | 0.121 | 0.68 |
| 31 [Ag, A30] | 1.5 | 30 | 9.37 | 0.081 | 0.59 |
| 46 [Ag, A30] | 1.0 | 30 | 9.91 | 0.032 | 0.35 |
| 94 [Ag, A30] | 0.5 | 30 | 9.4 | 0.009 | 0.20 |
| 469 [Ag, A30] | 0.1 | 30 | 8.8 | 0.000 | 0.04 |
| 45 [Cu]* | 2.0 | 100 | 13.429 | 0.0070 | 0.74 |
| 90 [Cu]* | 2.0 | 100 | 13.232 | 0.0072 | 1.52 |
| 135 [Cu]* | 2.0 | 100 | 13.636 | 0.0068 | 2.16 |

* indicates as-prepared films that have not been compacted.

**Discussion**

The thermal measurements show two notable features: (1) the A30 sample has a much lower thermal diffusivity, despite having a similar density, compared to the A60, and (2) compaction leads to a very significant decrease in thermal diffusivity in both of the Ag samples, and in the Cu sample. The former feature is likely the result of our choice of nanowires, specifically the diameter. While both nanowires are high purity and single crystal, and so should reflect the published values of bulk thermal diffusivity, the A30 sample has a diameter that is smaller than the mean free path of electrons in silver. This causes there to be additional scattering from the surface of the nanowire, suppressing the ability of the electrons to carry heat along the nanowire. By comparison, the A60 nanowires are larger than the electron mean free path, and so will carry heat further before scattering.

Addressing the latter, we consider the nanostructure of the aerogel. Generally, as a material is compacted, the thermal diffusivity is increased due to a corresponding increase in the number of conduction pathways through the material. However, for these aerogels, the conduction pathway is through the nanowires and across junctions. Compaction of the aerogel collapses voids within the material, but does not fundamentally change the network pathway that heat-carrying electrons must take – namely, along the wires and across the junctions. This behavior resembles a topology-preserving densification process, where large geometric compression collapses pore volume without fundamentally restructuring the interconnected conductive nanowire network. As a result, the Euclidean pathlength represented by the sample thickness and the effective conductive pathway through the nanowire network are distinct. Since the thermal diffusivity formalism assumes heat transport across a homogeneous continuum with a well-defined geometric thickness, it does not fully capture transport in these highly porous percolative nanowire networks. In the present system, the Euclidean thickness decreases substantially during compaction, while the effective conductive pathway through the interconnected nanowire network changes much more weakly. As a consequence, the fitted “b” coefficient likely provides a more direct representation of how thermal transport evolves during compaction than the calculated thermal diffusivity alone.

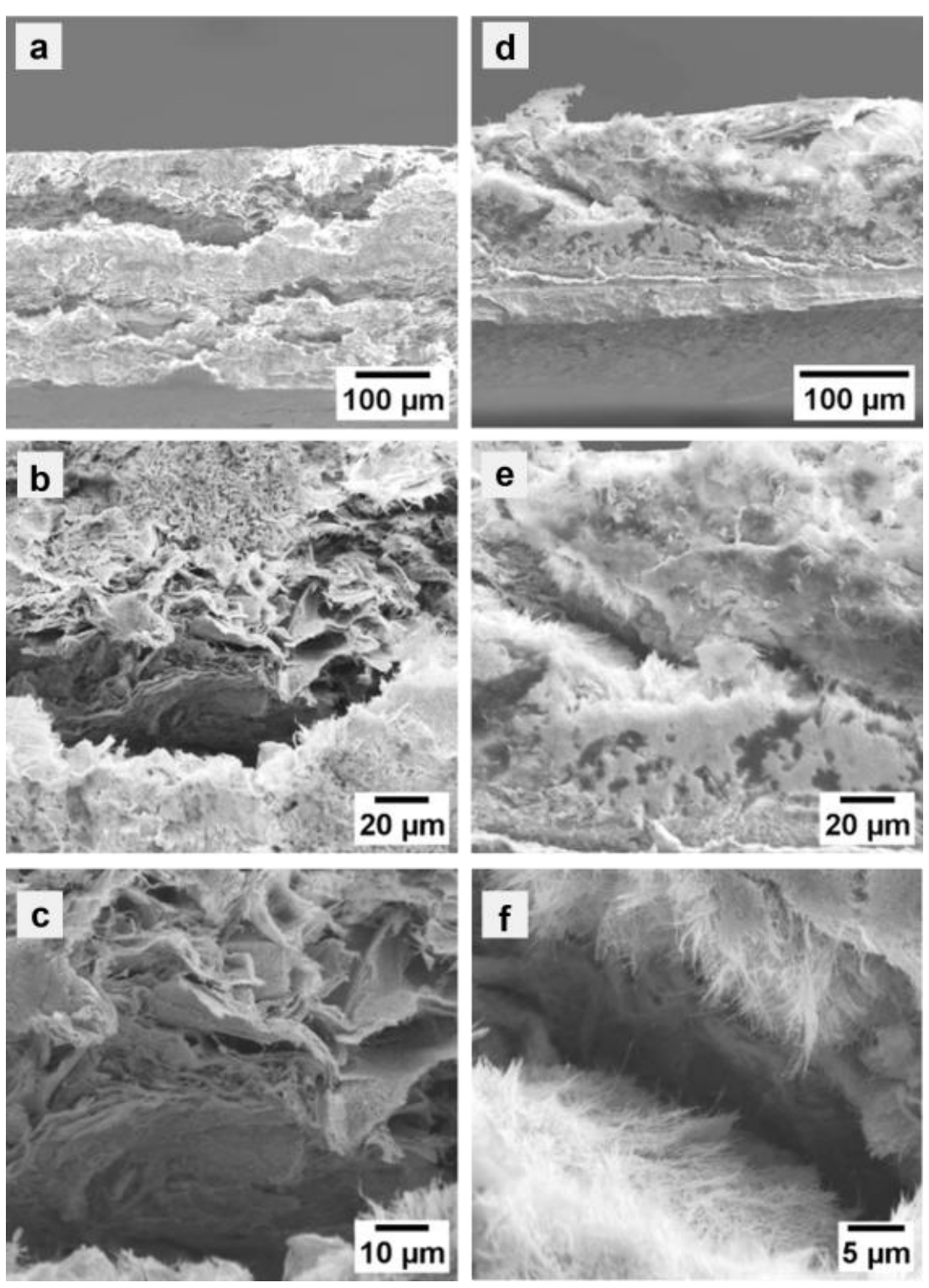

**Figure 5** Cross-section-view SEM images of compacted A30 aerogel (a, b, c) and compacted A60 aerogel (d, e, f).

To better represent the physical heat transport process, we instead consider the characteristic through-thickness thermal diffusion time, which can be calculated by $t \approx \frac{d^2}{\alpha} = \frac{b^2}{\pi}$. Plotting $\frac{b^2}{\pi}$, Fig. 4c, shows a decrease in the timescale of 22% for A30 and 72% for A60 during compaction. This is consistent with the interpretation above: in an absolute sense, thermal equilibration across the sample becomes faster as the aerogel is compacted, however the change is comparatively modest relative to the ≈300× decrease in the calculated thermal diffusivity. Conventional treatment using thermal diffusivity – and thermal conductivity, which is proportional to the diffusivity – is dominated by the Euclidean thickness. As a consequence the thermal diffusivity – which encodes the thickness-normalized rate of heat transfer – decreases substantially faster. Beyond the mesoscale Euclidean versus actual pathlength, other nanoscale factors such as bending or breaking of the wires and an increasing number of junctions also affects the thermal diffusivity. These factors are likely responsible for the 'modest' (4×) change in propagation time. Altogether, this is a lesson that these nanostructured materials should not be treated as a bulk average: the density increases by 30×, resulting in an unexpected decrease in the thermal diffusivity of >300× and thermal conductivity of 17×, meanwhile a heat wave moves through the material ≈4× faster than the uncompacted material. Functionally, this is a result with important implications for heat transfer in shields, such as on satellites, that even after significant compaction, the characteristic thermal transport timescale remains comparatively similar.

As a final note, the 30× increase in density with a corresponding 4× increase in thermal heat rate still suffers an order of magnitude difference. Inspecting the cross section of the material reveals that the compaction process significantly modified the microstructure, with large transverse voids forming within the material, **Figure 5**. As a result, thermal transport through the material is severely interrupted, however the transverse thermal conduction may be enhanced significantly more than the through-thickness values we report here, resulting in anisotropic transport.

## Conclusions

This work investigated thermal transport in metallic nanowire aerogels as a function of nanowire structure and sequential mechanical compaction. Nanoscale confinement within the nanowires was shown to strongly suppress thermal transport, likely due to enhanced surface scattering once the wire diameter became comparable to or smaller than the electron mean free path. Mechanical compaction reduced the calculated thermal diffusivity by nearly 300×, however this decrease was shown to be dominated by the thickness dependence embedded within the thermal diffusivity formalism. In contrast, the measured thermal transit time through the material decreased by only ≈4×, indicating that compaction primarily collapses pore volume while comparatively weakly modifying the underlying conductive nanowire network. This behavior resembles a topology-preserving densification process, where large geometric compression does not proportionally alter the effective percolative transport pathway through the material. Altogether, these results demonstrate that highly porous nanowire aerogels cannot always be treated as homogeneous bulk media, as the macroscopic geometry and the conductive network evolve differently during densification. Instead, the collective thermal behavior is strongly governed by the underlying nanostructure and network connectivity.

## Acknowledgements

This work was supported by the Department of Energy under awards SC0021344 and SC0026383, and DARPA under award FA8650-21-2-7125.